\documentclass[preprint,12pt]{elsarticle}

\usepackage{amssymb}

\usepackage[nodots]{numcompress}

\usepackage{subfigure}

\newcommand{\arthsix}{$^{36}$Ar}
\newcommand{\ar}{$^{39}$Ar}
\newcommand{\pot}{$^{39}$K}
\newcommand{\arfor}{$^{40}$Ar}
\newcommand{\potfor}{$^{40}$K}
\newcommand{\kr}{$^{85}$Kr}
\newcommand{\keVee}{keV}

\journal{Astroparticle Physics}

\begin{document}

\begin{frontmatter}



\title{A Study of the Residual \ar\ Content in Argon
from Underground Sources}

\author[PU]{J.~Xu}
\author[PU]{F.~Calaprice \corref{cor1}}
\ead{frankc@princeton.edu}
\author[PU]{C.~Galbiati}
\author[LNGS]{A.~Goretti}
\author[PU]{G.~Guray}
\author[PU]{T.~Hohman \fnref{PUEng}}
\author[PU]{D.~Holtz \fnref{JHU}}
\author[LNGS]{A.~Ianni}
\author[LNGS]{M.~Laubenstein}
\author[PU]{B.~Loer \fnref{FNAL}}
\author[TU]{C.~Love \fnref{Former_Name}}
\author[TU]{C.J.~Martoff}
\author[PU]{D.~Montanari \fnref{FNAL}}
\author[HU]{S.~Mukhopadhyay}
\author[PU]{A.~Nelson}
\author[VT]{S.D.~Rountree}
\author[VT]{R.B.~Vogelaar}
\author[PU]{A.~Wright}

\address[PU]{Department of Physics, Princeton University, Princeton NJ 08544, USA}
\address[LNGS]{INFN Laboratori Nazionali del Gran Sasso, SS 17 bis Km 18\_910, 067010 Assergi (AQ), Italy}
\address[TU]{Physics Department, Temple University, Philadelphia, PA 19122, USA}
\address[HU]{Department of Earth and Planetary Sciences, Harvard University, Cambridge, MA 02138, USA}
\address[VT]{Physics Department, Virginia Polytechnic Institute and State University, Blacksburg, VA 24061, USA}

\cortext[cor1]{Corresponding author}

\fntext[PUEng]{Current address: Department of Mechanical and Aerospace Engineering, Princeton University, Princeton NJ 08544, USA}
\fntext[JHU]{Current address: Department of Physics and Astronomy, The Johns Hopkins University, Baltimore, MD 21218, USA}
\fntext[FNAL]{Current Address: Fermi National Accelerator Laboratory, Batavia, IL 60510, USA}
\fntext[Former_Name]{Formerly C. Martin}

\begin{abstract}

The discovery of argon from underground sources with significantly less \ar\ than atmospheric argon was
an important step in the development of direct-detection
dark matter experiments using argon as the active target. 
We report on the design and operation of a low background detector with a
single phase liquid argon target that was built to study the \ar\ content of
the underground argon. Underground argon from the 
Kinder Morgan CO$_2$ plant in Cortez, Colorado was determined to have less than 0.65\% of the \ar\ activity in atmospheric argon. 
\end{abstract}


\end{frontmatter}


\section{Introduction}
\label{intro}

Argon is a bright scintillator that has attractive features for use in experiments attempting to directly detect Weakly Interacting Massive Particle (WIMP)-type dark matter.  Argon-based detectors can make use of the pulse shape discrimination (PSD) capability inherent to argon scintillation light to separate low energy nuclear recoil events, produced by WIMP-nuclear collisions, from electron recoils, produced by gamma ray or beta background.   The efficiency for rejecting electron recoil events can be very high, with demonstrated inefficiencies currently approaching $10^{-8}$~\cite{DEAP}.  Argon-based detectors can also be operated as ionization detectors, or as combined scintillation-ionization Time Projection Chambers (TPCs)~\cite{WARP_3kg_2008} with even more background suppression capabilities. The relatively low cost of argon technology also makes the deployment of ton-scale target masses feasible.

Argon-based direct detection dark matter searches, however,  must confront the presence of  intrinsic \ar, a beta emitter with an endpoint energy of 565\,keV.  \ar\ is present at an activity  of $\sim$1\,Bq/kg~\cite {Benetti_2007_39ar_activity} in commercial argon derived from the atmosphere.  

Boulay and Hime were the first to observe that the argon scintillation pulse shape discrimination could suppress the \ar\ background sufficiently to allow sensitive searches for dark matter~\cite{Boulay_Hime_2006}.   Even with pulse shape discrimination, however, the \ar\  activity in atmospheric argon limits the ultimate size of argon-based experiments and restricts their ability to probe very low energy events.

Most of the argon in the Earth's atmosphere was produced by electron capture of long-lived \potfor\ (\potfor\  + $e^{-} \rightarrow$  \arfor\  + $\nu$)  present in natural
potassium within the earth. The \ar\ activity in the atmosphere is
maintained by cosmogenic production through $^{40}$Ar(n,2n)$^{39}$Ar
and similar reactions. The half-life of \ar\ is  269\,yr and, as a result, the only practical way to reduce the amount of \ar\ in atmospheric argon is isotope separation. This is possible, but costly and time consuming for ton-scale argon detectors. 

Much of the radiogenic argon produced by the decay of \potfor\ is still present underground and can be found in gas wells at useful concentrations.  Even at modest depths ($\sim$100\,m) the overburden is sufficient to stop the hadronic component of cosmic ray showers that are responsible for most of the \ar\ production. Therefore, it might be expected that argon from underground sources would have less cosmogenic \ar\ radioactivity.

Unfortunately, as shown in studies by Loosli, Lehmann and Balderer~\cite{Loosli_1989} and Lehmann~\cite{Lehmann_1993}, \ar\  can be produced underground by a sequence of nuclear reactions wherein alpha particles from the decays of uranium and thorium and their progeny produce neutrons by the $(\alpha,n)$ reactions on light nuclei, and the neutrons then produce \ar\ by the $(n,p)$ reaction on stable potassium, \pot$(n,p)$\ar. 
For typical crustal composition with potassium concentrations of a few percent and uranium and thorium concentrations of a few ppm, the average \ar\ concentration in crustal argon is not expected to be significantly lower than that in atmospheric argon. 

However, because the uranium and thorium content of the crust varies, the \ar\ concentration in underground argon can vary from place to place~\cite{Lehmann_1993, Loosli_1983_39Ar_dating}; therefore, a search was undertaken for a large supply of underground argon low in \ar. An early promising measurement was made on argon that was extracted from the crude helium  in the National Helium Reserve~\cite{AcostaKane_2008}.  The crude helium contains argon and other volatile gasses that were separated in the production of natural gas.  Several firms, including Linde, have major facilities to purify the  crude helium to produce commercial helium.   Crude helium gas was collected at the Linde facility and sent to Princeton University where the argon was extracted and purified  with  a zeolite gas separation system~\cite{loer_thesis}. The argon was then sent to the laboratory of H. H. Loosli in Bern, Switzerland, for measurement of the \ar\ using low background gas counters.   Measurements showed no evidence for \ar, with an upper limit of 5\% of the level in atmospheric argon, as described in Ref.~\cite{AcostaKane_2008}.  

The result demonstrated that it is possible to locate underground argon with less \ar\ than is found  in atmospheric argon. The Linde site was especially attractive because processed argon could be obtained already separated from the natural gas in a relatively pure state.  For technical reasons, however, air is used in the helium production plant  to burn residual methane and in doing so, the underground argon is polluted with \ar\ from atmospheric argon.  The crude helium would have to be purified directly to avoid air contact.   Because of this complication, the National Helium Reserve was not pursued.  

Compared to the Earth's crust, which has uranium and thorium at ppm levels, the Earth's mantle is thought to contain these elements at the ppb level.  The radiogenic production of \ar\ depends on the product of concentration of the alpha emitters (uranium and thorium) and the concentration of  potassium, whereas radiogenic production of  \arfor\ depends only on the potassium concentration.  Thus, the ratio of \ar/\arfor\ depends on the concentration of the alpha emitters uranium and thorium, and should be approximately 1000 times lower for mantle gas compared to crustal gas. 
Based on helium, neon, argon, krypton and xenon measurements, the gas in the CO$_{2}$ gas fields in the US southwest is derived from the Earth's upper mantle~\cite{Staudacher_1987, Caffee_1999, Holland_Ballentine_2006}; these fields should therefore be a good source of argon low in \ar.

The zeolite system used to purify the argon extracted from the National Helium Reserve was upgraded to increase its capacity and was deployed at the Reliant Dry Ice facility on the Bravo Dome CO$_{2}$  gas field in Bueyeros, New Mexico. First samples showed no evidence for \ar,  but later measurements yielded evidence for \ar, possibly due to the decrease in well pressure, as the CO$_{2}$ deposit at that site was consumed.  The gas collection system was subsequently upgraded further~\cite{Babcock_2008} and moved to the Kinder Morgan CO$_2$ extraction facility in Cortez, Colorado. At this site, we are currently extracting argon at a rate of a few hundred grams per day~\cite{Babcock_2008}, and at the time of writing approximately 85\,kg of argon have been collected.  The results presented in this work are based on kg-sized samples of argon collected at Cortez in 2009 and 2010.  

The \ar\ measurement in reference~\cite{AcostaKane_2008} was made using a low mass gas detector,
and the sensitivity was limited by both statistics and background. 
Accelerator Mass Spectrometry (AMS) has also been used to measure \ar\ in argon with a 
sensitivity of 5\% of atmospheric~\cite{AMS1}, and before the measurements presented 
here was undertaken, an effort was made to improve the AMS methods to measure \ar\ 
to higher sensitivity. However, background from \pot\ could not be controlled 
well enough to improve on the previous results~\cite{AMS2}. 
Recently, Atomic Trap Trace Analysis (ATTA) has been reported to be 
able to perform an \ar/\arfor\ isotope analysis
below the atmospheric level, and future improvements may increase
the sensitivity to the level of 0.1\% of atmospheric~\cite{ATTA_2011}.

In this paper we describe a custom, low background, liquid argon calorimeter that was designed to
improve the sensitivity and the robustness of the \ar\ measurement.
The detector is described in Section~\ref{apparatus}. Using this
apparatus a new, more stringent upper limit of 0.65\% on the \ar\ activity in
argon extracted from the Cortez site relative to atmospheric argon has been determined, as described
in Section~\ref{results}. At this activity, experiments using
underground argon active volumes will significantly outperform those
with atmospheric argon, especially in their sensitivity to low mass WIMPs, and
\ar\ will not limit the size of dark matter experiments using underground
argon for the foreseeable future.

\section{Preparation of Underground Argon}	
 
 As described in Ref.~\cite{Babcock_2008}, the raw carbon dioxide gas at the Cortez site 
is processed by a vacuum-pressure-swing adsorption unit, 
producing a crude mixture of 70\% nitrogen, 27.5\% helium, and 2.5\% argon by volume.

The argon in this crude gas was further separated and purified at Princeton University for the measurement
presented here. A zeolite column, which contained 18\,kg of Zeochem's Zeox Z12-07 zeolite
in a 30\,L aluminum gas cylinder, was used to absorb the nitrogen content in the gas mixture. 
This zeolite, consisting of beads with diameters of between 0.4\,mm and 0.8\,mm, has a reported nitrogen adsorption capacity of at least 21\,mL/g at STP. The gas was fed into the zeolite column at a flow rate of 10\,L/min., producing an effluent stream of concentrated argon and helium, which was, in turn, passed through a 6L stainless-steel bottle filled with 2\,kg of OVC 4x8 charcoal manufactured by Calgon Carbon and immersed in liquid nitrogen. This cryogenic charcoal trap captured the argon while allowing the helium to escape.

During this process, a small fraction of argon was trapped in the zeolite column. This was recovered by purging the saturated zeolite column with helium; the desorbed gases were captured in another cryogenic charcoal trap for further reprocessing.

The argon-rich gas produced by this system has a nitrogen concentration less than 1\% by volume, as measured by an SRS UGA 300 residual gas analyzer. While preparing a second, more recent batch of underground argon from the Cortez site, we demonstrated that further iterations of this process can bring the nitrogen level below 0.1\%.

\begin{figure}[!htp]
\begin{center}
\includegraphics[angle=0, width=0.7\textwidth]{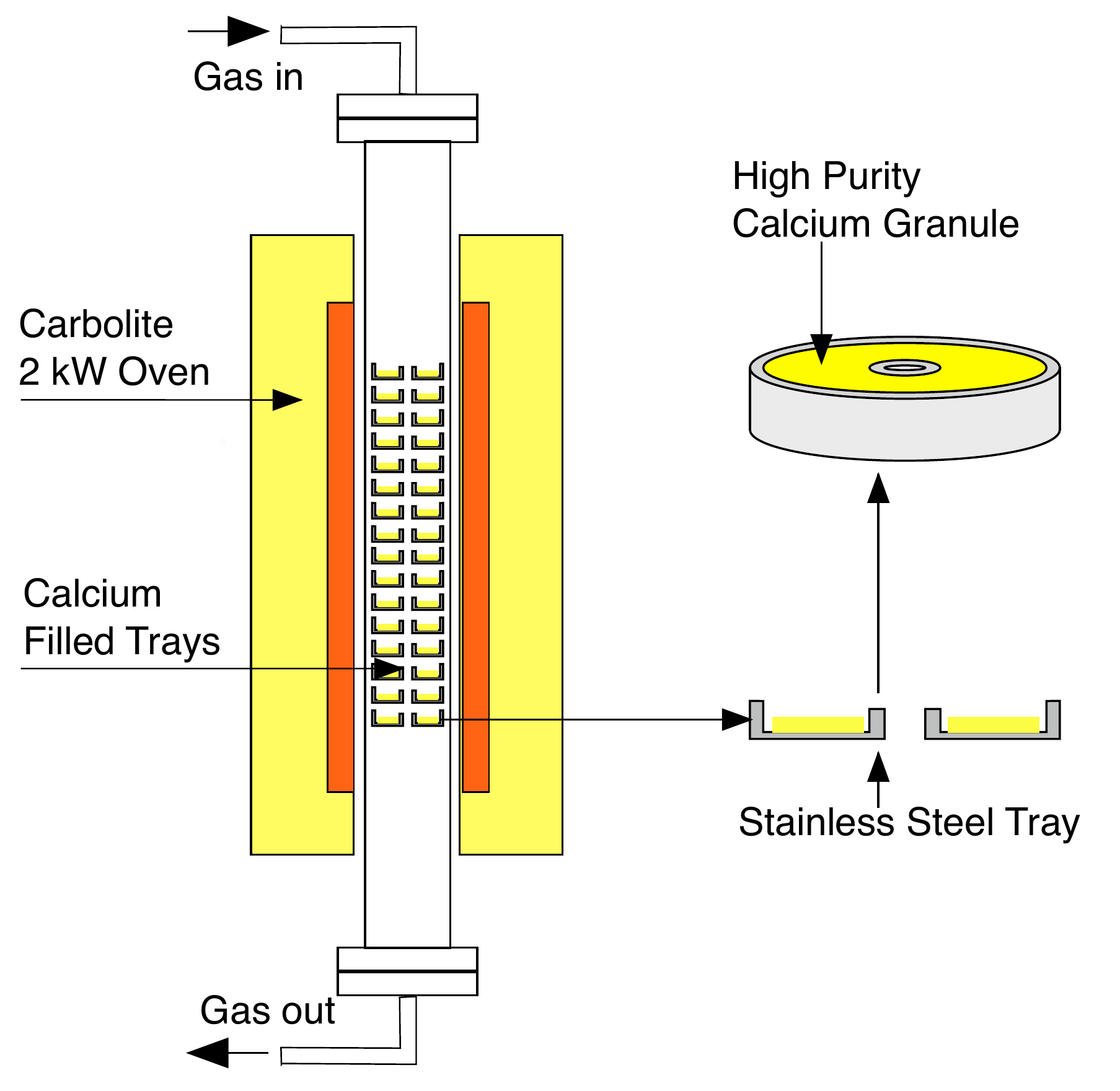}
\caption{The calcium trap used for argon purification. }
\label{fig:calcium_oven}
\end{center}
\end{figure}

Further treatment of the argon-rich gas involved the use of hot calcium to remove the remaining impurities. A calcium trap, illustrated in Fig.~\ref{fig:calcium_oven}, held trays that were filled with granules of $>$98.0\% purity calcium within a Conflat-sealed stainless-steel cylinder. The trap was heated by an external oven to 840 degrees Celsius, just below the melting temperature of calcium. The argon-rich gas was fed into the trap at 1\,L/min. and came into contact with the hot calcium, which was reacted with the impurities in the stream. The stable solids produced by these reactions, including calcium nitride, remained in the trays, while argon, as a noble gas, passed through the trap and was collected. The trap could be reactivated, as necessary, by simply replacing the calcium in the trays.

The specific underground argon sample used in this measurement had been stored on surface for about 3 years before the measurement, which is about 1\% of the \ar\ half life. Considering the equilibrium between \ar\ decay
and \ar\ production in the atmosphere, the generation of \ar\ in the sample during the storage period is a potential concern.  Fortunately, however, the cosmogenic neutron flux at surface is a few orders of magnitude lower than the average value in the atmosphere~\cite{Terrestrial_CR}, suggesting that \ar\ production at surface should be suppressed by a similar factor. A calculation using the COSMO code~\cite{cosmo} confirmed that the equilibrium \ar\ activity in argon stored on surface is only 1.7\,mBq/kg, so after three years of surface exposure the activity will be $\ll$1\,mBq/kg, well below the expected detection limit of the experiment.

\section{Apparatus}
\label{apparatus}

\begin{figure}[!htp]
\begin{center}
\includegraphics[angle=0, width=0.99\textwidth]{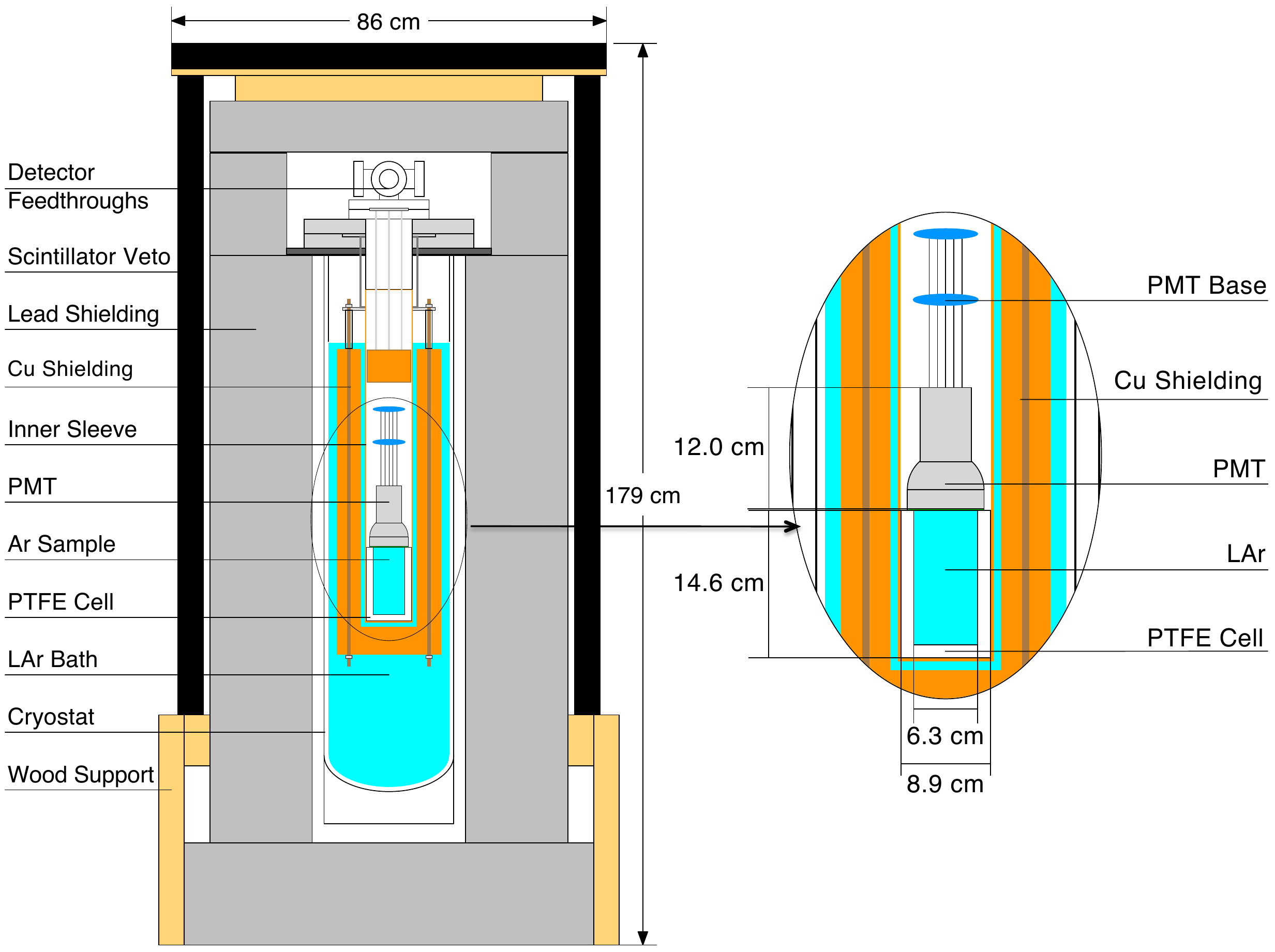}
\caption{A schematic diagram of the low background detector.}
\label{fig:detector}
\end{center}
\end{figure}

The low background detector, which is illustrated in
Fig.~\ref{fig:detector}, contains 0.56\,kg of liquid argon as
the active volume in a 2.5 inch inner diameter, 5.25 inch high ``cup'' made of highly
crystaline PTFE. The inner surfaces of the PTFE container and the fused silica top window
were coated with 300\,$\mu$g/cm$^2$ of p-Terphenyl to shift the wavelength
of the ultraviolet argon scintillation light into the visible range. The photons
were detected by a low background 3 inch photomultiplier tube (PMT) above the cup.
The cup and PMT were enclosed within a 3.5 inch diameter copper and
stainless steel ``sleeve, '' which was sealed at the top by a Conflat
flange. The sides and bottom of the sleeve were surrounded by 2 inch of oxygen-free high conductivity
(OFHC) copper shielding; a loose-fitting 2.5 inch thick copper plug could be inserted into the sleeve
above the PMT to complete the inner shielding. 

The use of a PMT with a low level of intrinsic radioactivity was important 
in obtaining low background operation. We
used a Hamamatsu R11065 metal bulb photomultiplier, which combines low
backgrounds with high quantum efficiency ($>$30\%) cryogenic operation. 
In order to reduce the amount of underground argon used in the experiment,
the PMT was not completely immersed in liquid argon, so a high-voltage divider
capable of operating in gaseous argon (which has a low dielectric
strength) was designed and built.

The inner detector, as well as the copper shielding around it,
were cooled by a bath of commercial liquid argon in an
open cryostat. The liquid level in the
cooling bath was maintained by a liquid level controller designed 
by Teragon Research (model LC10)
attached to a low pressure liquid argon supply dewar. 
The open cryostat sat
within, and was supported by, a lead castle that was 8 inch thick on the
sides, 12 inch thick on the bottom, and 4 inch thick on the top.
A muon veto system, composed of 2'' thick plastic scintillator
panels, was placed over the sides and top of the lead. 
 
\begin{figure}[!htp]
\begin{center}
\includegraphics[angle=0, width=0.75\textwidth]{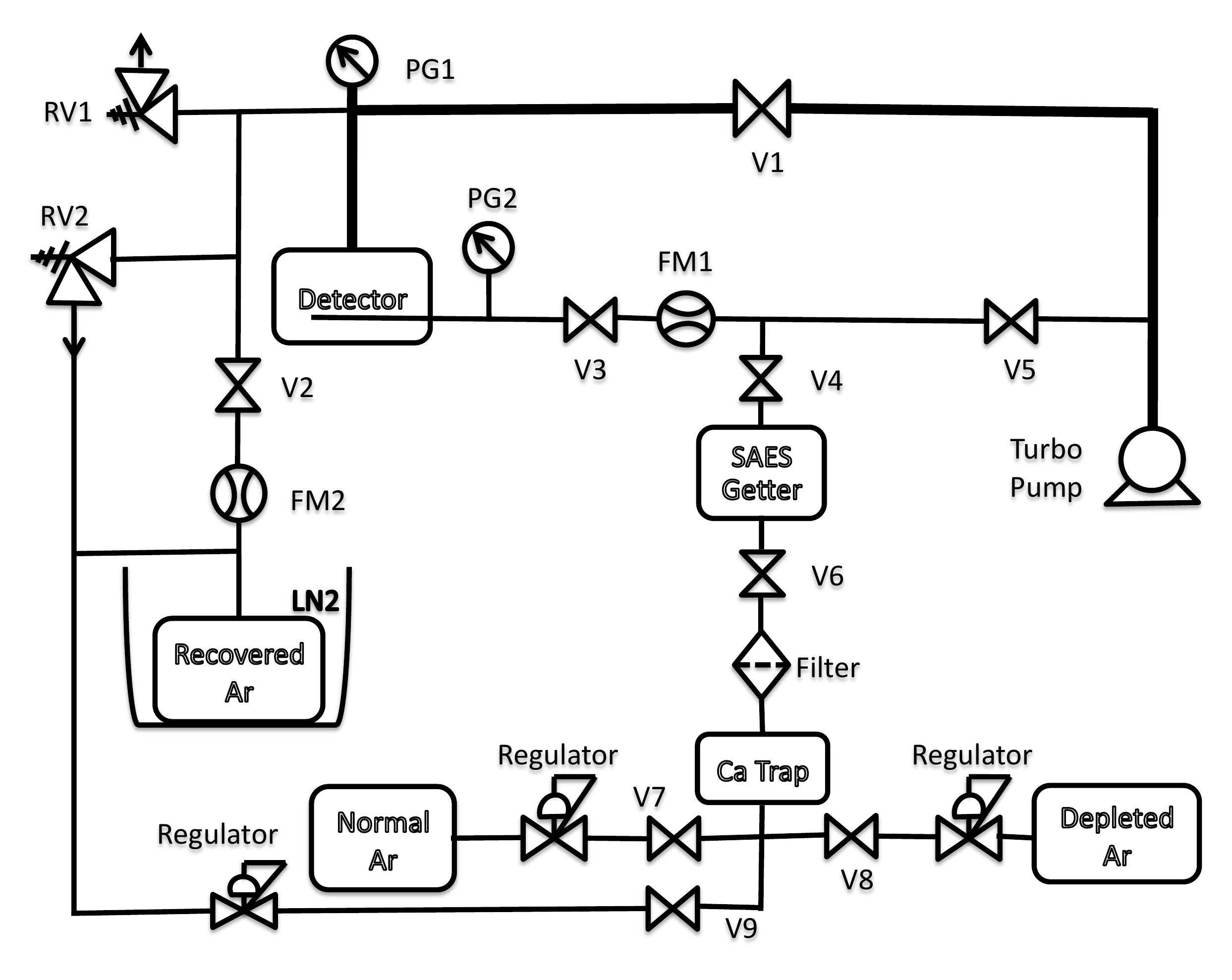}
\caption{The gas handling system for the low background argon detector.}
\label{fig:gashandling}
\end{center}
\end{figure}

The low background detector gas handling system, shown in Fig.~\ref{fig:gashandling},
 consisted of copper tubing with Swagelok
connections. The gas plumbing and the detector could be evacuated using
a turbo-molecular pump (Pfeiffer Vacuum HiCube), with typical vacuum levels of 10$^{-6}$\,mbar obtained. 
When the detector was being filled, the argon gas was purified using both the calcium trap 
described earlier and a Mono-Torr zirconium getter to remove impurities
such as water, oxygen, and nitrogen before being condensed in the
pre-cooled detector volume. To empty the detector, cooling was stopped and
the underground argon was recovered by condensing it in a charcoal-filled
aluminum gas cylinder partially immersed in a bath of liquid nitrogen.  The use of LN$_{2}$-cooled activated charcoal allowed for transfer and recovery of the argon without the use of mechanical compressors that could introduce impurities.

\begin{figure}[!htp]
\begin{center}
\includegraphics[angle=0, width=0.7\textwidth]{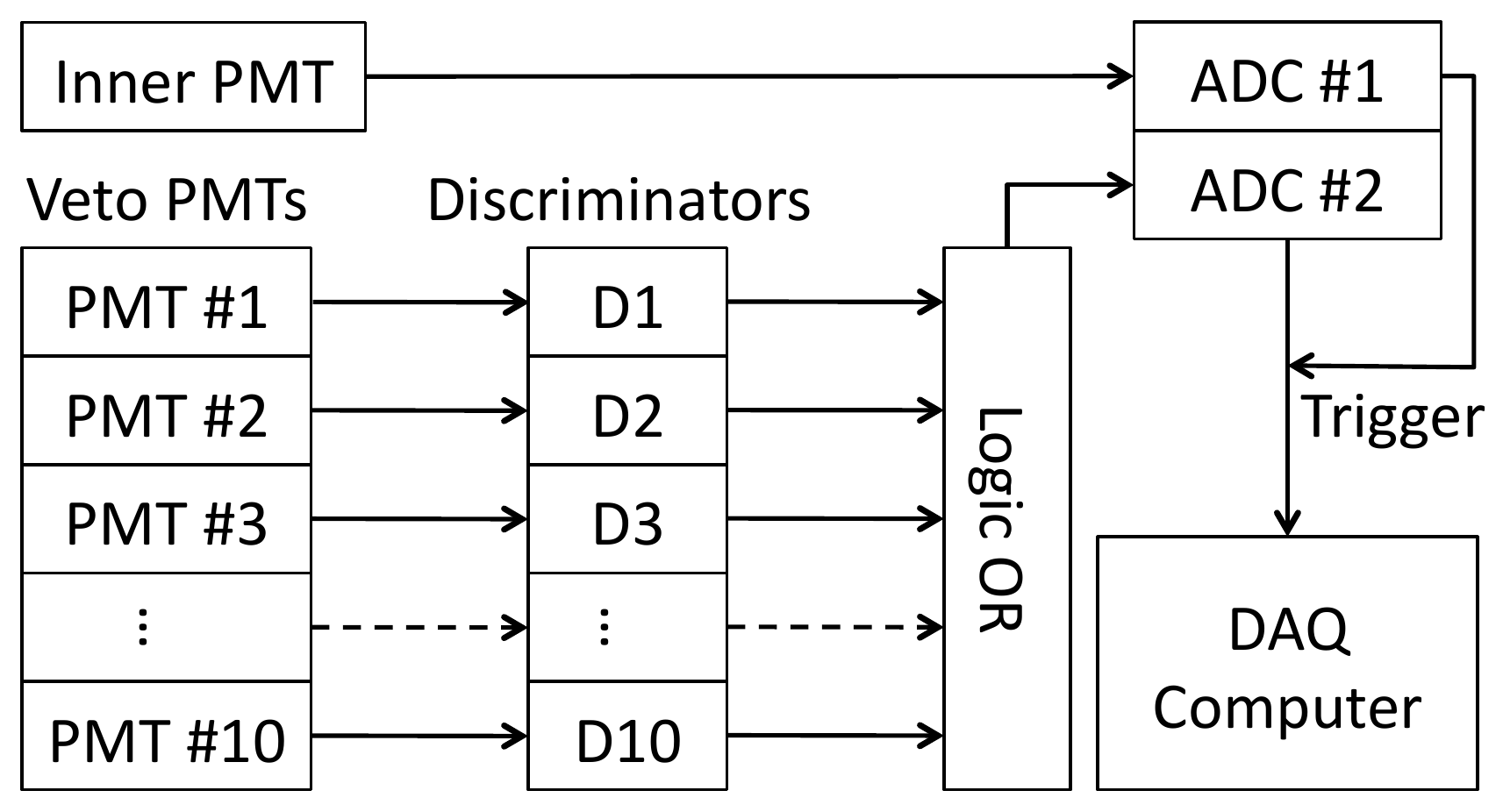}
\caption{The data acquisition system for the low background detector.}
\label{fig:daq}
\end{center}
\end{figure}

Data acquisition was carried out using 12\,bit, 250\,MHz digitizers (CAEN V1720). 
As illustrated in Fig.~\ref{fig:daq}, the output of the inner detector PMT was digitized directly, while
the signals from the ten PMTs of the muon veto system were processed by
LeCroy 623B amplitude discriminators 
to produce 50\,ns logic pulses which were summed and fed
into a second digitizer channel. Readout of both digitizer
channels was triggered by an amplitude discriminator
on the inner PMT channel. For each event, 15\,$\mu$s of data were
recorded, beginning 5\,$\mu$s before the trigger.

\section{Results for $^{39}$Ar}
\label{results}

\subsection{Data Collection}

The detector was operated in two main campaigns, one at surface in a
basement laboratory at Princeton University, and one underground in
the Kimballton Underground Research Facility (KURF) in 
Virginia. In both campaigns, the detector was operated with samples of
both atmospheric and underground argon. 

\begin{figure}[!htp]
\begin{center}
\includegraphics[angle=0, width=0.65\textwidth]{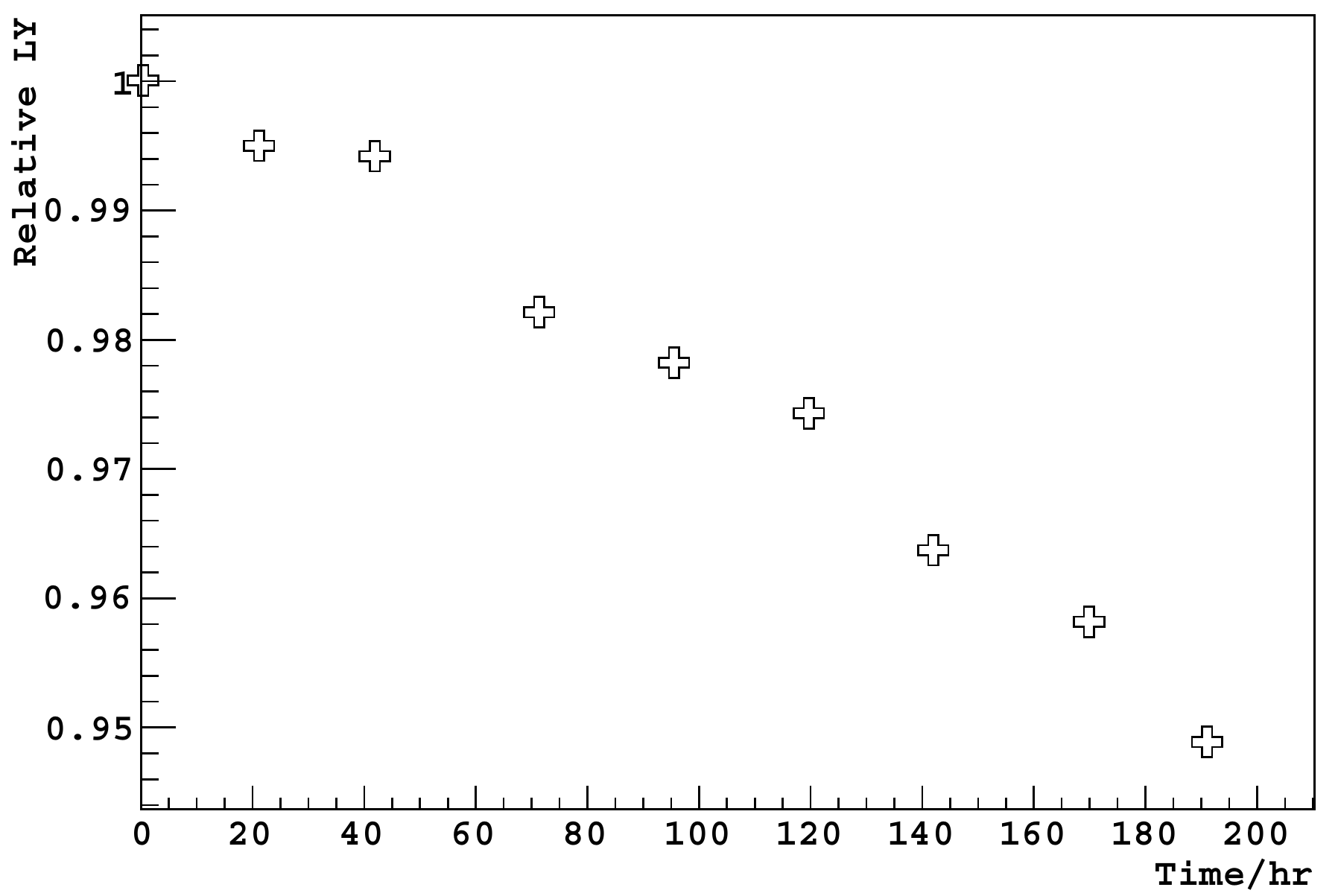}
\caption{The light yield degradation with time.}
\label{fig:lydegrade}
\end{center}
\end{figure}

The energy response of the detector was studied using a  $^{137}$Cs source
that could be introduced between the lead shielding of the detector and the cryostat
containing the bath of atmospheric argon. In addition, during atmospheric argon runs the energy
scale of the detector could be calibrated using the \ar\ decay spectrum: the two methods agreed to better than 1\%. 
Throughout this paper, energy is presented  in electron equivalent units, 
assuming a linear scintillation response scaled from the $^{137}$Cs calibration data; 
i.e., all events approximate minimum ionizing particles. 
While not true for nuclear recoil events and low-energy events~\cite{quenching}, 
the assumption holds well for the particles and energy range of interest for this analysis.

The light yield of the inner detector was evaluated 
using the mean charge of the single photoelectron (p.e.) peak as a reference.
In the runs at Princeton typical light yields of 6 - 7\,p.e./\keVee\ were 
achieved\footnote{The light yield could be further improved by increasing the photocathode 
coverage. A value of 9\,p.e./\keVee\ was observed with the DarkSide-10 
detector~\cite{DS10_LY}, with PMTs installed both on the top and on the bottom of the sensitive volume.},
while the light yield in the KURF runs was lower, at $\sim$5\,p.e./\keVee,
possibly due to the exposure of inner detector components to 
mine air during detector assembly.
As shown in Fig.~\ref{fig:lydegrade},  the light yield decreased slowly
 with time  ($<$ 1\% per day) as impurities released from 
 inner detector components accumulated.
However, given the relatively short duration of each measurement period 
(typically one week), in-run purification was not implemented in this detector.
The drift in the light yield was monitored with calibration data, 
and corrections were applied in the analysis.

\begin{figure}[!htp]
\begin{center}
\includegraphics[angle=0, width=0.75\textwidth]{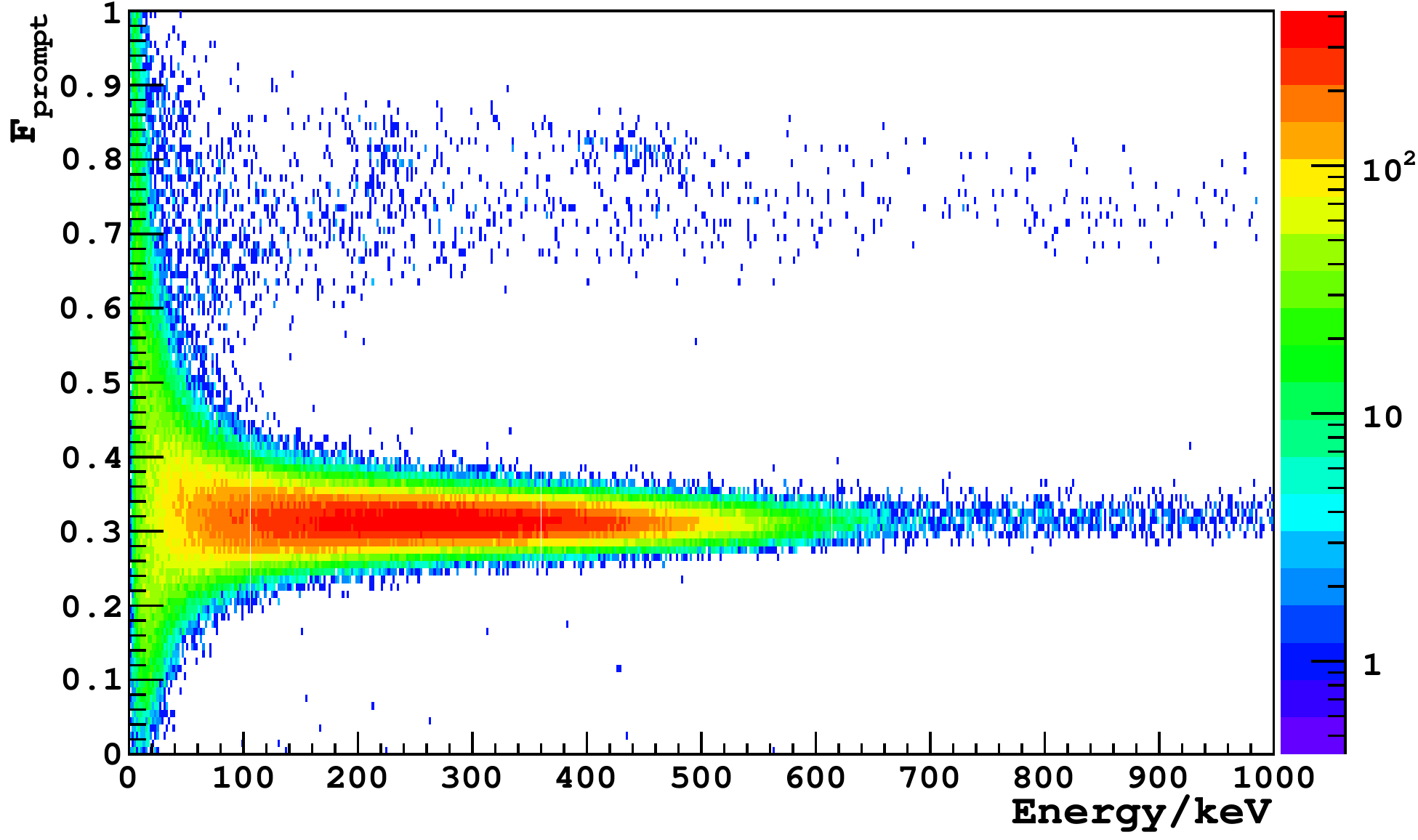}
\caption{F$_{\rm prompt}$ vs. energy in atmospheric argon data. Electron recoil events
form the bottom band with F$_{\rm prompt}$ $\sim$0.3, and nuclear recoil events form 
the top band with F$_{\rm prompt}$ $\sim$0.75.}
\label{fig:fpvse}
\end{center}
\end{figure}

The time profile of liquid argon scintillation pulses depends strongly on the density of ionization
produced by the exciting particle,
and thus the pulse shape of the scintillation signal can be used
to separate electron recoil events and nuclear recoil events with very high efficiency.
As shown in Fig.~\ref{fig:fpvse}, the fraction of light in an argon scintillation pulse that arrives
within the first 90\,ns (F$_{\rm prompt}$) 
is distinctly different in electron recoil events and nuclear recoil events.
A cut on F$_{\rm prompt}$ was used to select the electron recoil events for 
the \ar\ analysis, which was the only explicit cut applied to the data at KURF.
The data acceptance was greater than 99\% for any energy window above 50\,\keVee,
with almost perfect rejection of nuclear recoils.

 \begin{figure}[!htp]
\begin{center}
\includegraphics[angle=0, width=0.75\textwidth]{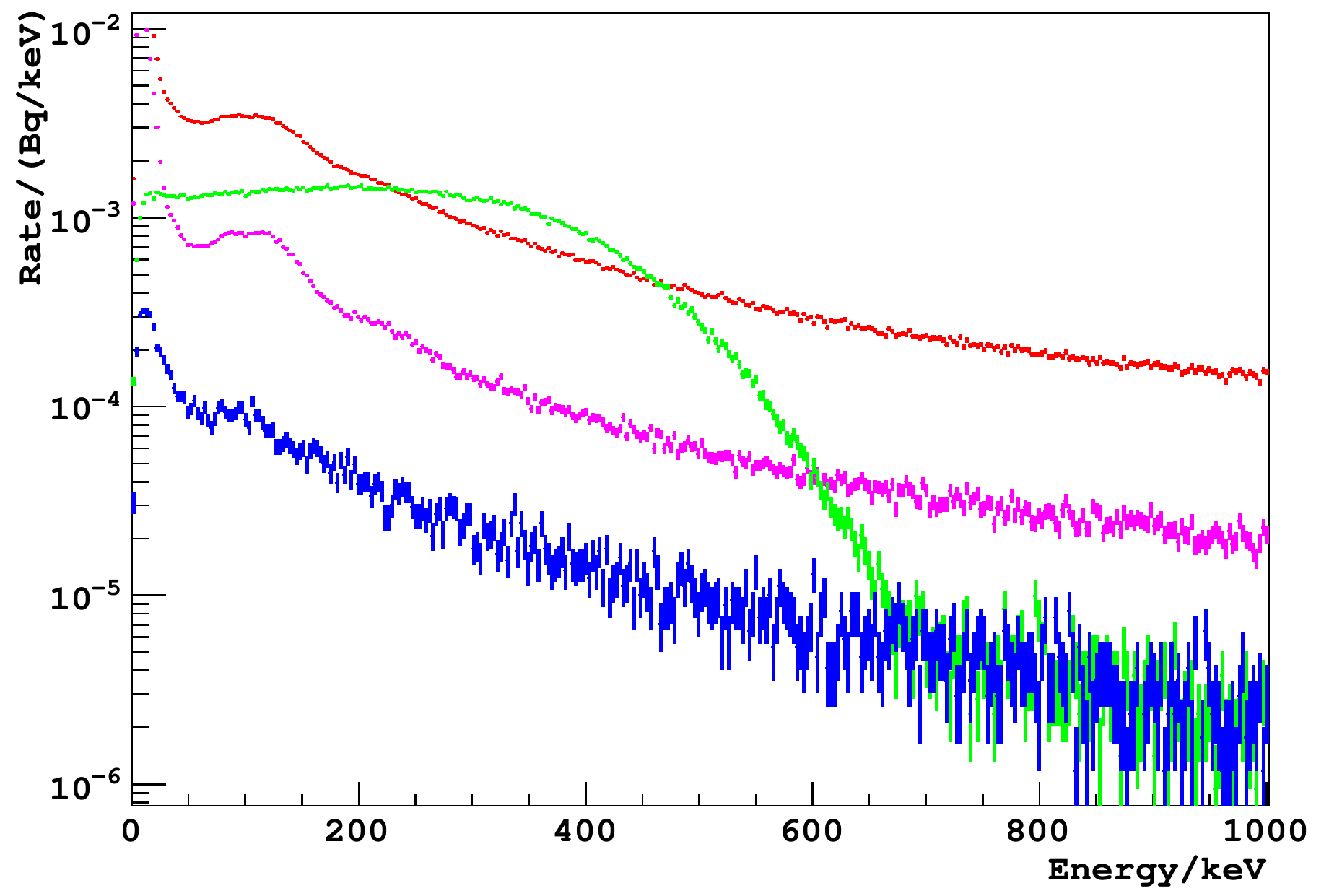}
\caption{The energy spectra recorded in the argon detector under different conditions.
Red: underground argon data at surface; 
purple: underground argon data at surface with an active cosmic ray veto;
blue: underground argon data at KURF;
green: atmospheric argon data at KURF.}
\label{fig:spectra}
\end{center}
\end{figure}

Fig.~\ref{fig:spectra} shows the energy spectra taken with the underground argon in different conditions.
The atmospheric argon spectrum at KURF is also plotted for reference. 
The muon veto system reduced the background rate by a factor of 5 at surface, 
and relocating the detector to KURF ($\sim$1450\,meters of water equivalent depth) 
caused the event rate to drop by another factor of 5. 
The cosmic ray muon rate at KURF was measured, using two
of the muon veto panels stacked horizontally, to be $\sim$1\,$\mu$/m$^2$/min, which is
approximately 10,000 times lower than that at surface. Thus the muon veto cut had no noticeable
effect on the underground data, and it was disabled to avoid unnecessary dead time. 
A residual event rate of 20\,mBq was achieved in the 50-800\,\keVee\ \ar\ window
in the measurement of underground argon at KURF.

\subsection{Rate Analysis}

\begin{figure}[!htp]
\begin{center}
\includegraphics[angle=0, width=0.75\textwidth]{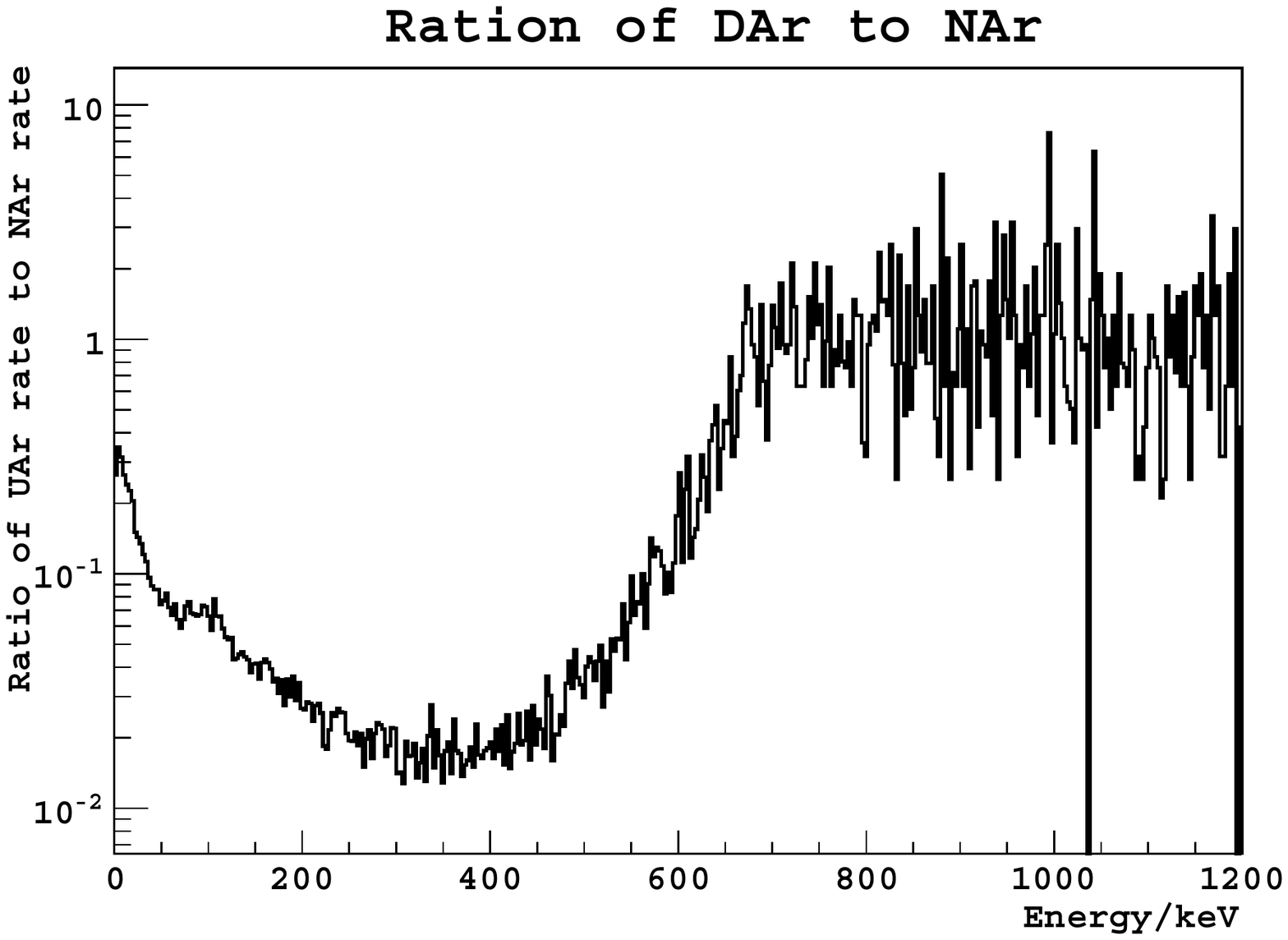}
\caption{The ratio of the event rate in underground argon to that in atmospheric argon.}
\label{fig:dnratio}
\end{center}
\end{figure}

Approximately 100\,kg$\cdot$hr each of underground argon data and atmospheric argon data
were collected at KURF and were used for this analysis. 
A conservative upper limit on the \ar\ content in the underground argon was obtained
by attributing all of the activity in the underground argon sample to \ar.
Fig.~\ref{fig:dnratio} shows the ratio of the event rate in underground argon to that in atmospheric
argon as a function of energy, and it indicates that the best \ar\ limit can be obtained in the 300 -
400\,\keVee\ window. The residual event rate in this energy window in the underground argon data 
after applying the PSD cut was  (1.87 $\pm$ 0.06)\,mBq, which is  (1.71 $\pm$ 0.05)\% of the corresponding event rate in the
atmospheric argon spectrum. This ratio can be taken as a first estimate of the relative \ar\ activity
in underground argon compared to atmospheric; however, given that the measured event rate in underground argon  also includes background events, this estimate must be taken only as a conservative upper limit.

\subsection{Background Subtraction}

A more stringent limit on the \ar\ activity was obtained by subtracting 
the estimated contributions from known background radioactivity from the 
observed events rate. 

Three Hamamatsu R11065 PMTs and the electronic components for one PMT base were sent to
the Gran Sasso Low Background Counting Facility~\cite{LNGS_counting} for radioactivity measurement,
the results from which are shown in Tab.~\ref{tab:measure_bkg}.
A model of the low background detector was constructed using the Geant4 Monte Carlo
simulation package~\cite{geant4}, which allowed us to
extrapolate the measured radioactivities to expected background rates
in our experiment.

\begin{table}[h!]
\begin{center}
\begin{tabular}{ c | c | c | c | c }
Decay & Measurement & PMT    & Base     & Cu\\
Chain & Point                 & (mBq)  & (mBq)   & (mBq/kg) \\
\hline
\hline
$^{232}$Th & $^{228}$Ra & 6 $\pm$ 1 & 41$\pm$2.8 & - \\
\cline{2-5}
& $^{228}$Th & 6 $\pm$ 1 & 45$\pm$4.7 & -\\
\hline
$^{238}$U & $^{234}$Th & 190 $\pm$ 40 &25 $\pm$ 3.7 & -\\
\cline{2-5}
& $^{234m}$Pa & 80 $\pm$ 40 & $<$ 149 &-\\
\cline{2-5}
& $^{226}$Ra & 18 $\pm$ 1.2 &32 $\pm$ 1.9 & - \\
\hline
$^{235}$U & $^{235}$U & 8 $\pm$ 2 &1.4 $\pm$ 0.4 & -\\
\hline
$^{40}$K & $^{40}$K & 79 $\pm$ 10 & 65 $\pm$ 9.3& - \\
\hline
$^{60}$Co & $^{60}$Co & 8.8 $\pm$ 0.8 & $<$ 1.2 & 2.1 $\pm$ 0.19\\
\hline
$^{57}$Co & $^{57}$Co & - & - & 1.8 $\pm$ 0.4\\
\hline
$^{58}$Co & $^{58}$Co & - & -  & 1.7 $\pm$ 0.09\\
\hline
$^{56}$Co &$^{56}$Co  & - & -  & 0.2 $\pm$ 0.03\\
\hline
\end{tabular}
\caption{The major radioactive isotopes in detector components.}
\label{tab:measure_bkg}
\end{center}
\end{table}

The cosmogenic activity in the OFHC copper shielding was also considered.
The specific copper sample in the experiment was not counted; however, equilibrium cosmogenic
copper activation values, as reported in Ref.~\cite{matthias} and as listed in 
Tab.~\ref{tab:measure_bkg}, were used. The copper measured in the reference was exposed to cosmic rays at LNGS, Italy (altitude 985\,m above sea level), and the activities were scaled down by a factor of 2.1 to reflect the lower cosmic ray flux at sea level, as suggested in the paper. The exposure time of our specific copper sample was also taken into account. Our Geant4 simulation was used to translate the copper activities into detector background rates. A 25\% uncertainty was added in the later analysis to account for the lack of knowledge of copper exposure history, the variations in the cosmic ray flux with latitude, and the (small) overburdens covering our copper during exposure.

While the backgrounds from the detector materials described above are expected to be identical in the measurements of atmospheric and underground argon, the presence of krypton in the argon could produce a background signal that is different between the two measurements. Krypton contains  \kr, an anthropogenic isotope with a beta decay endpoint energy of
687\,keV, very close to that of \ar. The level of krypton in the underground argon is expected to be extremely low
because no efficient underground krypton production mechanism is known,
and, in any case, \kr\ in the underground argon would be conservatively considered to be \ar\ in the analysis.
\kr\ in the atmospheric argon reference sample, on the other hand,
could artificially lower the apparent \ar\ activity ratio between the underground and atmospheric argon samples.
The krypton level in the high purity atmospheric argon purchased as our standard was measured
by the manufacturer to be less than 40\,ppb. With a typical \kr/Kr ratio of 15\,ppt, the \kr\ decay rate in the reference 
sample would be less than 1.8\% of the \ar\ decay rate. A systematic uncertainty was added 
in the analysis to account for this potential effect.

Besides the intrinsic background in the detector, it was discovered that a weak $^{252}$Cf
source used by another experiment at KURF contributed non-negligible background to 
the argon detector, even though it was stored 45\,ft away.
Removing the $^{252}$Cf source reduced the event rate in the 300-400\,\keVee\ window by 40\%,
but only 3 hours of $^{252}$Cf-free data were collected because the detector was scheduled
to shut down on the same day the source effect was discovered. As a result, the \ar\ limit 
determined from the data sample without neutron contamination was statistically limited.
The statistics could be improved by choosing a larger analysis energy window,
but this required knowledge of the neutron background spectrum.
Therefore, another run with the $^{252}$Cf source right beside the detector was taken, 
which constituted the neutron calibration spectrum, to study the neutron-induced events.

\begin{figure}[!htp]
\begin{center}
\includegraphics[angle=0, width=0.75\textwidth]{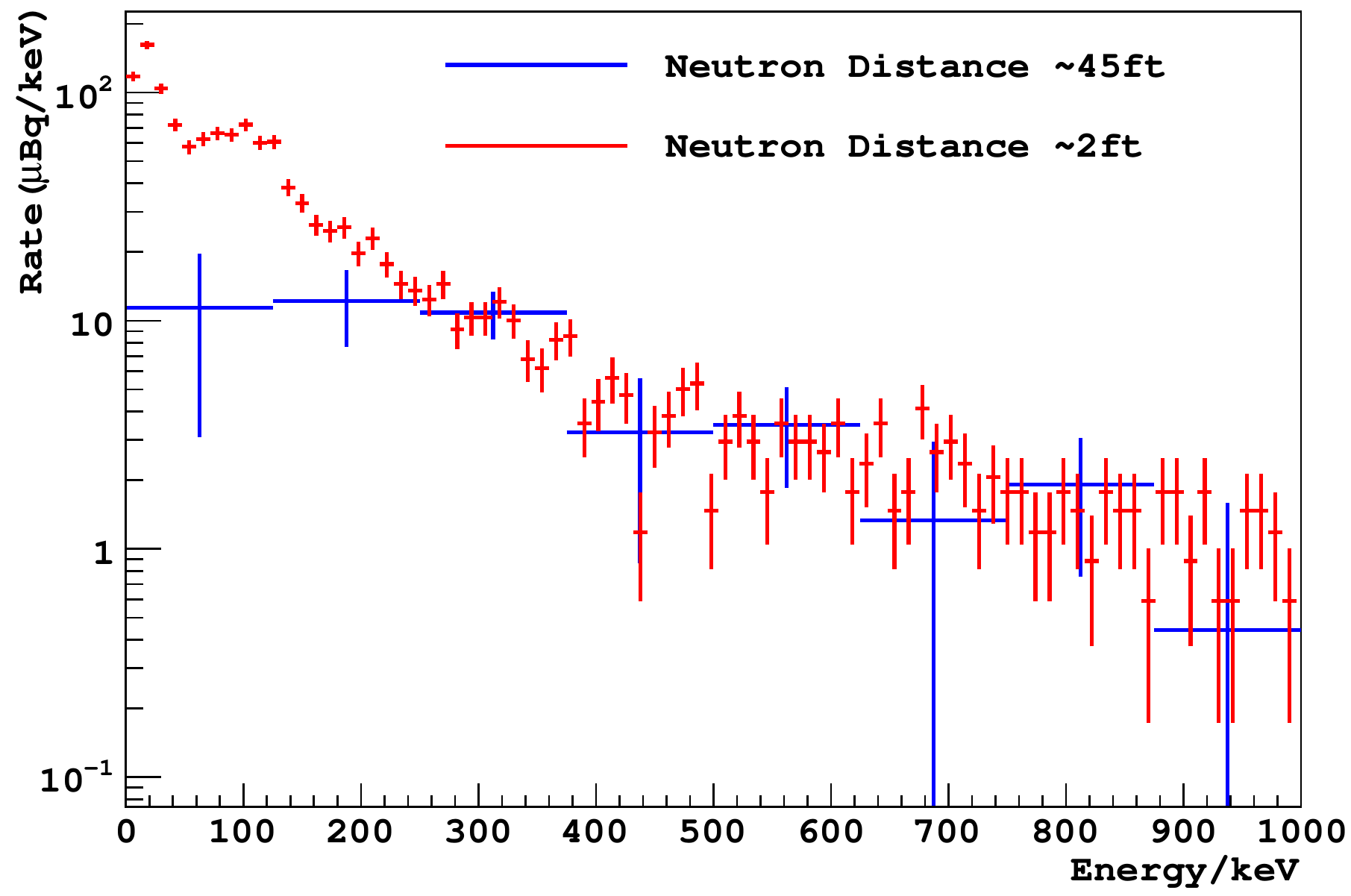}
\caption{The spectra of electron recoil events induced by the $^{252}$Cf neutron source 
at different distances from the argon detector. 
The red spectrum was scaled down to match the rate of the blue one above 250\,\keVee.}
\label{fig:cf252}
\end{center}
\end{figure}

Fig.~\ref{fig:cf252} shows the neutron spectrum acquired with the source right beside
the detector, overlaid with the low statistics spectrum produced by subtracting
the livetime-normalized energy spectrum from the 3 hours of data with no neutron source from
the corresponding spectrum with the $^{252}$Cf  source in its storage location.
The shapes of the two spectra are statistically consistent above 250\,\keVee, where the events
are believed to come from gamma ray emission following neutron capture 
in different detector components.

The discrepancy visible between the two spectra in Fig.~\ref{fig:cf252} below 250\,\keVee\ 
is likely due to the inelastic scattering of energetic neutrons 
from $^{19}$F nuclei in the PTFE, which produces gamma rays at 109.9\,\keVee\  and 197\,\keVee.
These gamma rays were also observed in the measurement on surface, where high energy
cosmogenic neutrons are prevalent, as can be seen in Fig.~\ref{fig:spectra}.
We believe the inelastic neutron scattering was not as significant 
when the $^{252}$Cf source was at its storage location
because the emitted neutrons scattered to lower energy before reaching the detector,
and were hence less likely to excite the $^{19}$F nuclei.

Because the energy distribution of neutron capture gamma rays is not expected to depend 
significantly on neutron energy, we assume that the spectrum above 250\,\keVee\ 
is the same for all source positions.
We therefore estimate the contamination rate from the neutron source in normal data runs 
by fitting the neutron calibration spectrum to the subtracted neutron spectrum
above 250\,\keVee, as is shown in Fig.~\ref{fig:cf252}, and using the
calibration spectrum to interpolate to the 300-400\,\keVee\ energy
range with minimal loss in statistical precision.

\begin{table}[h!]
\begin{center}
\begin{tabular}{ c  | c | c | c | c  }
\hline
Source & $^{252}$Cf &  PMT & Base &  Copper\\
\hline
\hline
Rate (mBq)& $0.82 \pm 0.16$ & $0.29 \pm 0.08$ & $0.07 \pm 0.02$ & $0.36 \pm 0.11$\\
\hline
\end{tabular}
\caption{The expected background rate in 300 - 400\,\keVee\ from different sources.}
\label{tab:bkgsum}
\end{center}
\end{table}

A summary of the expected background contributions in the 300 - 400\,\keVee\ 
energy region of the low background detector measurement is shown in Tab.~\ref{tab:bkgsum}. 
To be conservative, the contributions from detector components
whose radioactivities were not specifically measured are not included in the background estimation.
This includes, for example, the long lived radioisotopes in the OFHC copper shielding,
as well as any activity in the PTFE container. At typical radioactivity levels reported by 
other authors, these contributions would be small.

\begin{figure}[h!]
\begin{center}
\includegraphics[angle=0, width=0.75\textwidth]{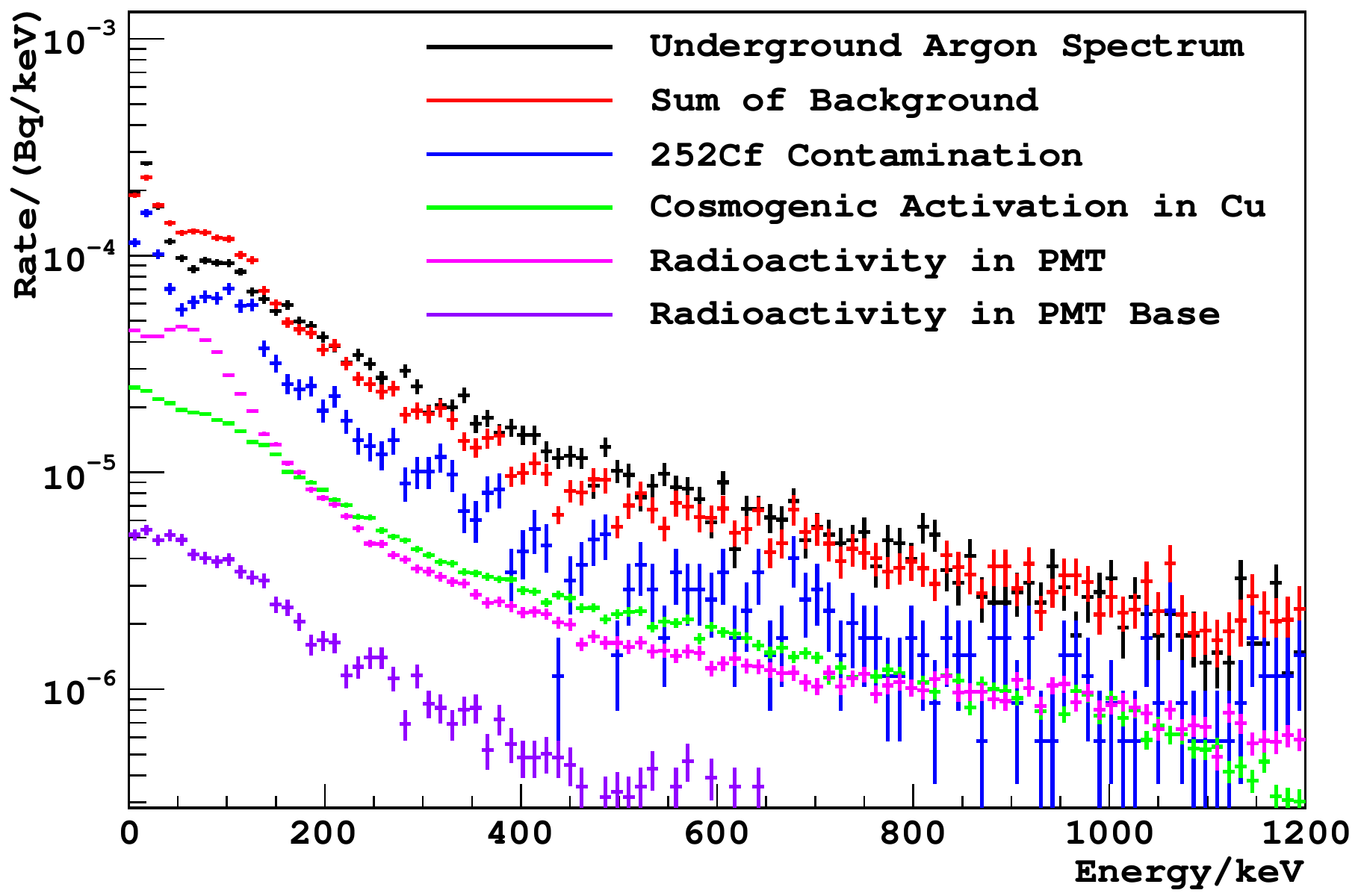}
\caption{The background spectra expected from the different identified background radioactivities compared to the observed spectrum.}
\label{fig:summary}
\end{center}
\end{figure}

In addition to the statistical uncertainties in the activity measurements listed in Tab.~\ref{tab:measure_bkg}, the PMT and base background estimates in Tab.~\ref{tab:bkgsum}.  include a 25\% uncertainty to account for the unknown spatial distribution of the radioactive contaminants within the components, and a similar 25\% uncertainty was assigned to the copper estimate due to factors discussed earlier. Additionally, we include a 5\% uncertainty to account for imperfections in the Monte Carlo geometry for all radioactivity simulations. 

A comparison of the predicted background spectra with our data  is shown in Fig.~\ref{fig:summary}; we believe the excess of events in the summed background spectrum around 100\,\keVee\ is due to neutron inelastic scattering in the neutron calibration data, as discussed above.

The known background components account for 80\% of 
the activity in the underground argon spectrum in 300 - 400\,\keVee. 
After subtracting the estimated backgrounds, a residual event rate of 
(0.32 $\pm$ 0.23)\,mBq remains in that energy region, as shown in Tab.~\ref{tab:subtracted}.
From this we deduce a 95\% C.L. upper limit of 0.65\% on the
\ar\ activity in underground argon relative to that in atmospheric argon. 

\begin{table}[h!]
\begin{center}
\begin{tabular}{  c | c  }
\hline
& Rate/mBq, (300, 400)\,keV \\
\hline
\hline
Natural Ar  (NAr) & $108.78\pm 0.39$\\
\hline
Underground Ar (UAr) & $1.87 \pm 0.06$\\
\hline
Estimated Background & $1.54 \pm 0.22$\\
\hline
$^{85}$Kr Background & $<1.83$\\
\hline
NAr, Background Subtracted  & $107.2 \pm 1.9$\\
\hline
UAr, Background Subtracted & $0.32 \pm 0.23$\\
\hline
\end{tabular}
\caption{A summary of the background subtraction analysis. The entire 
upper limit \kr\ rate is taken as an uncertainty in the background subtracted NAr rate. 
To convert these rates into activities per unit mass, an argon active mass of 0.56 $\pm$
0.03\,kg can be used.}
\label{tab:subtracted}
\end{center}
\end{table}

\section{Stable isotope ratios}
Atmospheric argon has a ratio of \arfor/\arthsix\ of 295.5~\cite{Ar36_ratio}, whereas Mid Ocean Ridge Basalts, which were derived by partial melting of the upper mantle, have measured \arfor/\arthsix\  ratios of 296 to 40,000~\cite{Sarda_1985, Marty_1997}. The large argon isotopic variability in the basalts reflects mixing between the upper mantle argon, which has a \arfor/\arthsix\ value of 41,000, and atmospheric argon introduced into the basalts upon eruption onto the seafloor. By comparison, the \arfor/\arthsix\ ratio for argon from the Doe Canyon CO$_{2}$ gas well was measured to be 13,300, supporting the hypothesis that a large fraction of the Cortez gas is derived from the mantle. By contrast, the argon from the National Helium Reserve had a measured \arfor/\arthsix\ ratio of 1,640, which is between atmospheric and mantle values. The low \ar\ value for the National Helium Reserve compared to expected levels from the crust, and the intermediate value of the \arfor/\arthsix\  ratio may suggest that the source of the gas is deeper than the  crust.  The argon from the National Helium Reserve was separated  from natural gas that is expected to be biotic, and therefore crustal in origin; however, noble gases in crustal fluids  can show a mantle signature, particularly in regions that may be undergoing tectonic extension~\cite{ONions_1988}.

\section{Future Work}

The dominant backgrounds in this measurement were identified
to be: 1) the unexpected $^{252}$Cf source, 2) the radioactivity
in the PMT and base, and  3) the cosmogenic activity in the copper
shielding.  By removing 1) and replacing the PMT with a new version with even
lower background, we estimate that the background event rate in
the \ar\ measurement can be reduced by at least a factor of two. A
second measurement campaign, implementing these improvements and
making use of a more recent batch of underground argon from
the Cortez extraction facility, is planned for the early summer of
2012. In the longer term, a more significant upgrade in which
the passive shielding of the current experiment is replaced by
an active veto composed of organic liquid scintillator is being
considered.

\section{Conclusion}
\label{discussion}

A single phase argon detector with a total electron recoil background rate of
$<4$\,mHz/kg in the 300 - 400\,\keVee\ energy window when operated in
an underground laboratory has been constructed and used to investigate
the \ar\ activity in argon from underground sources. We report no observation of \ar\ in argon gas
extracted from the Doe Canyon field in Cortez, Colorado, setting an upper
limit of 0.65\% on the \ar\ activity relative to atmospheric argon. This limit is almost 
10 times lower than earlier results~\cite{AcostaKane_2008} and, to the best of our knowledge, 
demonstrates the highest sensitivity to \ar\ in argon yet obtained.
This result also represents an important milestone in argon-based direct detection dark matter
searches, as at this activity multi-ton two-phase detectors can be
operated without pileup. The large reduction in
\ar\ activity will also allow current- and next-generation experiments
using underground argon to operate with lower threshold energies than
experiments using atmospheric argon. This will improve the sensitivity
of the underground argon experiments, especially in the low mass WIMP region. 

\section{Acknowledgments}

We acknowledge the hospitality of the Kimballton Underground Research
Facility, and thank the management and staff at Lhoist North
America - Kimballton for their valuable support and assistance. 
We thank Martin Cassidy and Martin Schoell for their invaluable help in identifying suitable sampling locations in New Mexico and Colorado. We thank the management and staff of Kinder Morgan for their support and assistance in hosting us at the Doe Canyon facility in Cortez, Colorado. We gratefully acknowledge the loan of the plastic scintillator muon veto
and electronics from Fermi National Accelerator Laboratory and thank Stephen Pordes for his helpful contributions. We thank the Princeton Plasma Physics Laboratory for the use of some additional lead shielding.
This work was supported by NSF grants PHY0704220 and PHY0957083. AW acknowledges support from the Princeton PFEP program. We acknowledge valuable discussions with, and support from, our colleagues in the DarkSide collaboration.





\bibliographystyle{model1a-num-names}
\bibliography{bibliography}







\end{document}